\documentclass[aps,prl,twocolumn,superscriptaddress]{revtex4}
\usepackage{epsfig}
\usepackage{graphics}
\begin{document}
\title{Measurement of the $\beta^+$ and orbital electron-capture decay rates in
fully-ionized, hydrogen-like, and helium-like $^{140}$Pr ions}
\author{Yu.A.~Litvinov}%
%\email{y.litvinov@gsi.de}%
\affiliation{Gesellschaft f\"ur Schwerionenforschung GSI,
Planckstra{\ss}e 1, 64291
Darmstadt, Germany}%
\affiliation{Justus Liebig Universit{\"a}t Giessen,
Heinrich-Buff-Ring 16, 35392 Giessen, Germany}%
\author{F.~Bosch}%
\affiliation{Gesellschaft f\"ur Schwerionenforschung GSI,
Planckstra{\ss}e 1, 64291 Darmstadt, Germany}
\author{H.~Geissel}%
\affiliation{Gesellschaft f\"ur Schwerionenforschung GSI,
Planckstra{\ss}e 1, 64291 Darmstadt, Germany} \affiliation{Justus
Liebig Universit{\"a}t Giessen, Heinrich-Buff-Ring 16, 35392
Giessen, Germany}
\author{J.~Kurcewicz}%
\affiliation{Warsaw University, Hoza 69, 00-681 Warszawa, Poland}
\author{Z.~Patyk}%
\affiliation{Soltan Institute for Nuclear Studies, Hoza 69, 00681
Warsaw, Poland}
\author{N.~Winckler}%
\affiliation{Gesellschaft f\"ur Schwerionenforschung GSI,
Planckstra{\ss}e 1, 64291
Darmstadt, Germany}%
\affiliation{Justus Liebig Universit{\"a}t Giessen,
Heinrich-Buff-Ring 16, 35392 Giessen, Germany}%
\author{L.~Batist}%
\affiliation{St. Petersburg Nuclear Physics Institute,
188300 Gatchina, Russia}%
\author{K.~Beckert}%
\affiliation{Gesellschaft f\"ur Schwerionenforschung GSI,
Planckstra{\ss}e 1, 64291 Darmstadt, Germany}
\author{D.~Boutin}%
\affiliation{Justus Liebig Universit{\"a}t Giessen,
Heinrich-Buff-Ring 16, 35392 Giessen, Germany}
\author{C.~Brandau}%
\affiliation{Gesellschaft f\"ur Schwerionenforschung GSI,
Planckstra{\ss}e 1, 64291 Darmstadt, Germany}
\author{L.~Chen}%
\affiliation{Justus Liebig Universit{\"a}t Giessen,
Heinrich-Buff-Ring 16, 35392 Giessen, Germany}
\author{C.~Dimopoulou}%
\affiliation{Gesellschaft f\"ur Schwerionenforschung GSI,
Planckstra{\ss}e 1, 64291 Darmstadt, Germany}
\author{B.~Fabian}%
\affiliation{Justus Liebig Universit{\"a}t Giessen,
Heinrich-Buff-Ring 16, 35392 Giessen, Germany}
\author{T.~Faestermann}%
\affiliation{Technische Universit{\"a}t M{\"u}nchen, 85748
Garching, Germany}
\author{A.~Fragner}%
\affiliation{Stefan Meyer Institut f{\"u}r subatomare Physik, 1090
Vienna, Austria}
\author{L.~Grigorenko}%
\affiliation{Joint Institut for Nuclear Research, 141980 Dubna,
Russia}
\author{E.~Haettner}%
\affiliation{Justus Liebig Universit{\"a}t Giessen,
Heinrich-Buff-Ring 16, 35392 Giessen, Germany}
\author{S.~Hess}%
\affiliation{Gesellschaft f\"ur Schwerionenforschung GSI,
Planckstra{\ss}e 1, 64291 Darmstadt, Germany}
\author{P.~Kienle}%
\affiliation{Technische Universit{\"a}t M{\"u}nchen, 85748
Garching, Germany} \affiliation{Stefan Meyer Institut f{\"u}r
subatomare Physik, 1090 Vienna, Austria}
\author{R.~Kn\"obel}%
\affiliation{Gesellschaft f\"ur Schwerionenforschung GSI,
Planckstra{\ss}e 1, 64291 Darmstadt, Germany} \affiliation{Justus
Liebig Universit{\"a}t Giessen, Heinrich-Buff-Ring 16, 35392
Giessen, Germany}
\author{C.~Kozhuharov}
\affiliation{Gesellschaft f\"ur Schwerionenforschung GSI,
Planckstra{\ss}e 1, 64291 Darmstadt, Germany}
\author{S.A.~Litvinov}%
\affiliation{Gesellschaft f\"ur Schwerionenforschung GSI,
Planckstra{\ss}e 1, 64291
Darmstadt, Germany}%
\affiliation{Justus Liebig Universit{\"a}t Giessen,
Heinrich-Buff-Ring 16, 35392 Giessen, Germany}%
\author{L.~Maier}%
\affiliation{Technische Universit{\"a}t M{\"u}nchen, 85748
Garching, Germany}
\author{M.~Mazzocco}%
\affiliation{Gesellschaft f\"ur Schwerionenforschung GSI,
Planckstra{\ss}e 1, 64291 Darmstadt, Germany}
\author{F.~Montes}%
\affiliation{Gesellschaft f\"ur Schwerionenforschung GSI,
Planckstra{\ss}e 1, 64291 Darmstadt, Germany}\affiliation{Michigan
State University, East Lansing, MI 48824-1321, U.S.A.}
\author{G.~M{\"u}nzenberg}%
\affiliation{Gesellschaft f\"ur Schwerionenforschung GSI,
Planckstra{\ss}e 1, 64291 Darmstadt, Germany} \affiliation{Manipal
Academy of Higher Education, Manipal, KA 576104, India}
\author{A.~Musumarra}
%INFN-Laboratori Nazionali del Sud, Via S.Sofia 44, I95123 Catania, Italy
%Dipartimento di Metodologie Fisiche e Chimiche per l'Ingegneria, Universita' di Catania, I95123 Catania, Italy
\affiliation{INFN-Laboratori Nazionali del Sud, I95123 Catania,
Italy}\affiliation{Universita$\acute{}$ di Catania, I95123
Catania, Italy}
\author{C.~Nociforo}%
\author{F.~Nolden}%
\affiliation{Gesellschaft f\"ur Schwerionenforschung GSI,
Planckstra{\ss}e 1, 64291 Darmstadt, Germany}
\author{M.~Pf\"utzner}%
\affiliation{Warsaw University, Hoza 69, 00-681 Warszawa, Poland}
\author{W.R.~Pla\ss}%
\affiliation{Justus Liebig Universit{\"a}t Giessen,
Heinrich-Buff-Ring 16, 35392 Giessen, Germany}
\author{A.~Prochazka}%
\affiliation{Gesellschaft f\"ur Schwerionenforschung GSI,
Planckstra{\ss}e 1, 64291 Darmstadt, Germany}
\author{R.~Reda}%
\affiliation{Stefan Meyer Institut f{\"u}r subatomare Physik, 1090
Vienna, Austria}
\author{R.~Reuschl}%
\affiliation{Gesellschaft f\"ur Schwerionenforschung GSI,
Planckstra{\ss}e 1, 64291 Darmstadt, Germany}
\author{C.~Scheidenberger}%
\affiliation{Gesellschaft f\"ur Schwerionenforschung GSI,
Planckstra{\ss}e 1, 64291
Darmstadt, Germany}%
\affiliation{Justus Liebig Universit{\"a}t Giessen,
Heinrich-Buff-Ring 16, 35392 Giessen, Germany}%
\author{M.~Steck}%
\author{T.~St\"ohlker}%
\affiliation{Gesellschaft f\"ur Schwerionenforschung GSI,
Planckstra{\ss}e 1, 64291 Darmstadt, Germany}
\author{S.~Torilov}%
\affiliation{St. Petersburg State University, 198504 St.
Petersburg, Russia}
\author{M.~Trassinelli}%
\affiliation{Gesellschaft f\"ur Schwerionenforschung GSI,
Planckstra{\ss}e 1, 64291 Darmstadt, Germany}
\author{B.~Sun}%
\affiliation{Gesellschaft f\"ur Schwerionenforschung GSI,
Planckstra{\ss}e 1, 64291 Darmstadt, Germany}\affiliation{Peking
University, Beijing 100871, China}
\author{H.~Weick}%
\author{M.~Winkler}%
\affiliation{Gesellschaft f\"ur Schwerionenforschung GSI,
Planckstra{\ss}e 1, 64291 Darmstadt, Germany}
\begin{abstract}
%
%The radioactive decay of nuclei, especially $\beta$-decay, is of
%utmost importance for the synthesis of the chemical elements by
%various processes in stars at high temperatures and, thus, at high
%ionization stages. With the advent of ion storage rings and traps
%we have the very first opportunity of studying the change of
%half-lives of radioactive nuclei with their degree of ionisation.

We report on the first measurement of the $\beta^+$- and orbital
electron capture decay rates of $^{140}$Pr nuclei with the most
simple electron configurations: bare nuclei, hydrogen-like and
helium-like ions.
%with one and two electrons in the K-orbit, respectively.
The measured electron capture decay constant of hydrogen-like
$^{140}$Pr$^{58+}$ ions is about 50\% larger than that of
helium-like $^{140}$Pr$^{57+}$ ions. Moreover, $^{140}$Pr ions
with one bound electron decay faster than neutral
$^{140}$Pr$^{0+}$ atoms with 59 electrons. To explain this
peculiar observation one has to take into account the conservation
of the total angular momentum, since only particular spin
orientations of the nucleus and of the captured electron can
contribute to the allowed decay.
%Besides its fundamental interest, these
%observations have profound impact on the understanding of beta
%decay of highly-charged ions in the stellar nucleosynthesis.
%
\end{abstract}
%\keywords{}
%
\maketitle
\par%
Various ways to influence nuclear decay rates have been tried by
scientists since radioactivity was discovered. Their motivation
reaches from the basic understanding of nuclear decay phenomena
and astrophysical reactions to applications like the transmutation
of nuclear waste. Small effects of up to a few percent have been
observed in atoms by changing the environmental parameters such as
pressure, temperature or electromagnetic fields
\cite{Emery,Ohtsuki}. These changes are mainly attributed to
modifications of the electron density at the nucleus. Significant
modifications of the electron conversion rate have been measured
when swift highly-ionized radioactive ions emerge from matter and
their nuclear decay is determined in flight~\cite{Ph-PRA,At-PRC}.
\par%
It has been predicted that the decay properties of highly-ionized
nuclides can be altered dramatically: decay modes known in neutral
atoms can become forbidden, new ones can be opened up. This can
have substantial impact on the nucleosynthesis in hot stellar
plasmas \cite{Bahcall,Takahashi}.
% new text
Seminal results on decay studies with selected highly-ionized ions
have been obtained with novel experimental tools using the
combination of high-energy accelerators, in-flight separators and
storage rings \cite{Geissel-PRL}.
% old text
%Based on novel experimental tools, decay studies with selected
%highly-ionized ions can be performed nowadays using the
%combination of high-energy accelerators, in-flight separators and
%storage rings \cite{Geissel-PRL}. Seminal results have been
%obtained meanwhile.
%
For example, the electron capture and electron conversion decays
become impossible in the absence of orbital electrons, i.e. in
fully-ionized atoms. Thus, the pure $\beta^+$-decay branch has
been measured in $^{52}$Fe$^{26+}$ ions \cite{Ir-PRL} and the
half-lives of isomeric states were found to be dramatically
prolonged \cite{Li-PLB}.
\par%
Bare $^{187}$Re$^{75+}$ ions decay, due to the new decay mode--the
bound-state $\beta$-decay--by nine orders of magnitude faster than
neutral $^{187}$Re atoms with a half-life of 42 Gyr \cite{Bo-PRL}.
Note that the couple $^{187}$Re/$^{187}$Os is used as a cosmic
clock. Bare $^{163}$Dy$^{66+}$ nuclei, being stable as neutral
atoms, become radioactive, thus allowing the s-process, the
astrophysical slow-neutron capture process of nucleosynthesis, to
branch \cite{Ju-PRL}. We note, that a simultaneous measurement of
$\beta$-decay to the continuum and bound states in
$^{207}$Tl$^{81+}$ ions has been performed recently \cite{Oh-PRL}.
\par%
In the present experiment the $\beta^+$ and orbital
electron-capture (EC) decays of bare nuclei and nuclei with one
and two bound electrons have been investigated. The EC decay rates
in hydrogen- and helium-like ions have been measured for the first
time.
%
%original paragraph
%In the present experiment, the $\beta^+$ and orbital
%electron-capture (EC) decay rates for bare nuclei and nuclei with
%only one and two bound electrons have been measured for the first
%time.
\par%
For this experiment we have selected the $^{140}$Pr (Z=59)
nucleus. The neutral atom decays with 99.4\% to the ground state
of $^{140}$Ce via a pure Gamow-Teller $\beta$-decay with a change
of the nuclear angular momentum by one unit ($\Delta I=1$) and no
parity change \cite{TOI}. The weak branches to the excited states
in $^{140}$Ce can be neglected in our context. A proton in
$^{140}$Pr can be converted into a neutron via a weak decay in two
ways, namely via the EC decay whereby a monochromatic
electron-neutrino is emitted
(p~+~e$^-$~$\rightarrow$~n~+~$\nu_e$), or via a three-body decay
in which the positron and the neutrino share the decay energy
(p~$\rightarrow$~n~+~e$^+$~+~$\nu_e$).
\par%
The experiment has been performed at the Gesellschaft f{\"u}r
Schwerionenforschung (GSI), Darmstadt, Germany, where the
combination of a heavy-ion synchrotron (SIS) \cite{SIS}, the
in-flight fragment separator FRS \cite{FRS} and the ion
storage-cooler ring ESR \cite{ESR} provides unique experimental
conditions for decay studies of bare and few-electron exotic
nuclei in an ultra-high vacuum ($\sim10^{-11}$~mbar). It is
possible to produce, separate and store exotic nuclei up to
uranium with a well-defined number of bound electrons
\cite{{Geissel-PRL},{Ir-PRL},{Li-PLB},{Bo-PRL},{Ju-PRL},{Oh-PRL}}.
Radioactive $^{140}$Pr ions have been produced via the projectile
fragmentation of $\sim$3$\cdot$10$^9$ $^{152}$Sm~ions/spill,
accelerated by the SIS to 508 MeV/u. A 1~g/cm$^2$ thick beryllium
target has been used. The fully-ionized, hydrogen- and helium-like
$^{140}$Pr ions were separated in-flight by a two-fold magnetic
rigidity analysis by means of the B$\rho$-$\Delta$E-B$\rho$
separation method \cite{FRS} in the FRS and subsequently injected
into the ESR. The flight time from the production target to the
storage ring was a few hundred nanoseconds. The ion-optical
settings of the FRS, the charge state distributions and the energy
degraders used in this experiment are described in details in Ref.
\cite{Li-AR}.
\par%
Stochastic \cite{No-NIMA} and electron cooling \cite{St-NIMA} were
applied to the $^{140}$Pr$^{59+}$, $^{140}$Pr$^{58+}$ and
$^{140}$Pr$^{57+}$ ions coasting in the ESR. The stochastic
cooling provides fast pre-cooling at a fixed fragment velocity,
corresponding to 400 MeV/u energy, thus reducing the overall
cooling time to about 2 seconds. The cooling forces all stored
ions to the same mean velocity and reduces the initial velocity
spread, caused by the fragmentation reaction, to $\delta v/v
\approx 5\cdot10^{-7}$.
\par%
%%%%%%%%%%%%%%%%%%%%%%%%%%%%%%%%%%%%%%%%%%%%%%%%%%%%%%%%%%%%%%%%%%%%%%%%%%%
\begin{figure}[t!]
\includegraphics*[width=8.0cm]{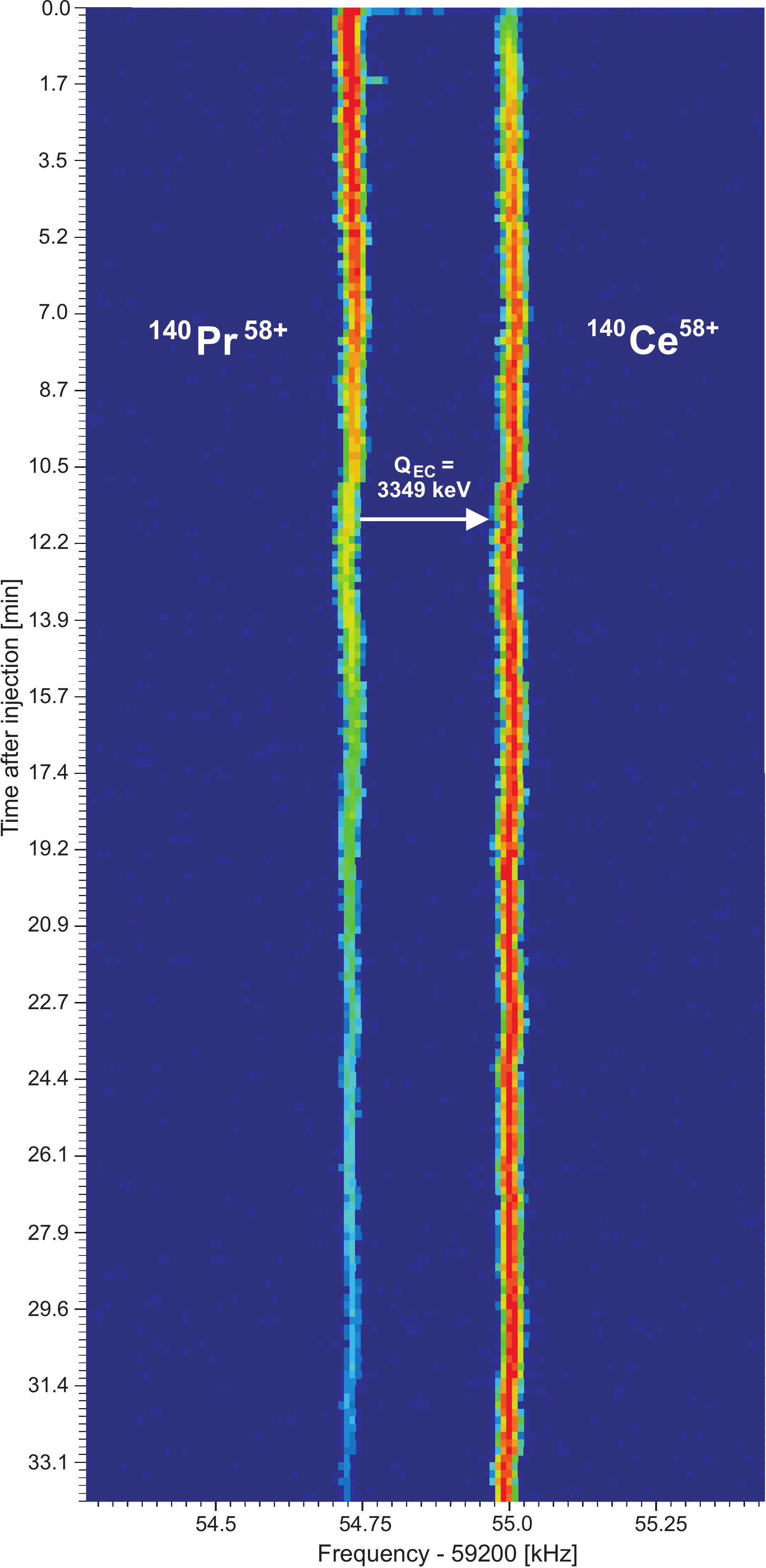}
\caption {Schottky frequency spectra at the 31$^{st}$ harmonics of
the revolution frequency taken subsequently as a function of time
(195 spectra \'a 10.5 sec). In the EC decay of hydrogen-like
$^{140}$Pr, the mass changes by 3.349 MeV/$c^2$ which leads to a
small change in the frequency ($\sim$270 Hz). The intensity of the
frequency lines is proportional to the number of stored ions. It
can be seen that the intensity of the line corresponding to the
parent ions $^{140}$Pr$^{58+}$ decreases in the course of time and
that the intensity of the line corresponding to the daughter ions
$^{140}$Ce$^{58+}$ increases.} \label{trace}
\end{figure}
%%%%%%%%%%%%%%%%%%%%%%%%%%%%%%%%%%%%%%%%%%%%%%%%%%%%%%%%%%%%%%%%%%%%%%%%%%%
The unambiguous identification of cooled $^{140}$Pr$^{59+}$,
$^{140}$Pr$^{58+}$ and $^{140}$Pr$^{57+}$ ions and their decay
products has been achieved exploiting the time-resolved Schottky
Mass Spectrometry \cite{{Li-NPA734},{Li-NPA756}}. The latter is
based on Schottky-noise spectroscopy \cite{Bo-SMS}, which is
widely used for non-destructive beam diagnostics in circular
accelerators and storage rings. The stored ions are circulating in
the ESR with revolution frequencies of about 2 MHz. At each turn
they induce mirror charges on two electrostatic pick-up
electrodes. Fast Fourier Transform of the amplified signals yields
the revolution frequency spectra, which provide information about
the mass-over-charge ratios of the ions. The area of the frequency
peaks is proportional to the number of stored ions, which is the
basis for lifetime measurements
\cite{{Ir-PRL},{Li-PLB},{Bo-PRL},{Ju-PRL},{Oh-PRL}}. The details
of the data acquisition system and of the data treatment can be
found in Ref. \cite{Li-NPA756} and references cited therein.
\par%
In the EC decay the atomic mass changes but the atomic charge
state is preserved. Therefore, this decay causes a sudden change
in the revolution frequency of about 270~Hz (31$^{st}$ harmonics).
An example for the $^{140}$Pr$^{58+}$ + e$^-$ $\rightarrow$
$^{140}$Ce$^{58+}$+ $\nu_e$ decay is illustrated in
Fig.~\ref{trace}, where 195 subsequent Schottky frequency spectra
are plotted as a water-flow diagram. Each spectrum is averaged
over 10.5 s. It can be seen in Fig.~\ref{trace} that the intensity
of the peak at lower revolution frequency--corresponding to the
parent ions $^{140}$Pr$^{58+}$--decreases steadily and that the
intensity of the peak at the higher frequency--corresponding to
the lighter daughter ions $^{140}$Ce$^{58+}$--increases.
%
%%%%%%%%%%%%%%%%%%%%%%%%%%%%%%%%%%%%%%%%%%%%%%%%%%%%%%%%%%%%%%%%%%%%%%%%%%%
\begin{figure}[t!]
\includegraphics*[width=8.0cm]{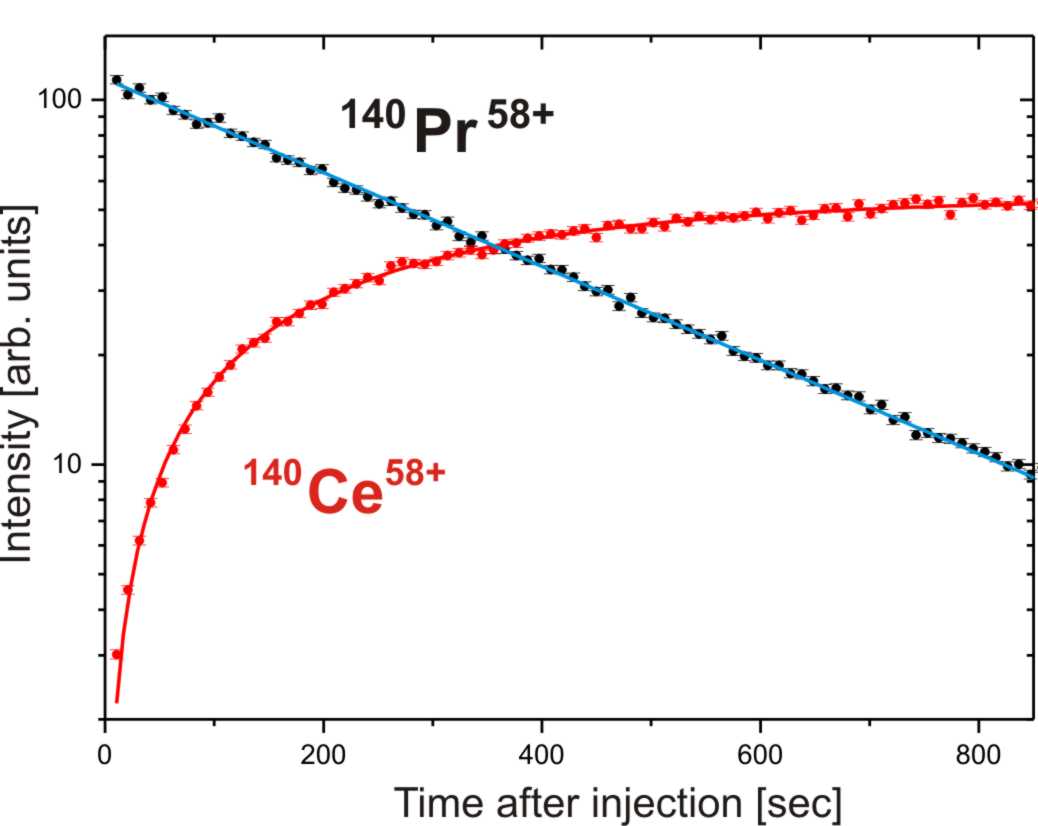}
\caption {Decay and growth curves of $^{140}$Pr$^{58+}$ and
$^{140}$Ce$^{58+}$ ions as a function of time. The data points are
shown in the laboratory frame and can be converted to the
rest-frame of the ions using the Lorentz factor $\gamma$ = 1.43.
The lines represent the fits according to equations (1) and (2).}
\label{decayc}
\end{figure}
%%%%%%%%%%%%%%%%%%%%%%%%%%%%%%%%%%%%%%%%%%%%%%%%%%%%%%%%%%%%%%%%%%%%%%%%%%%
Examples of the decay and growth curves are shown in
Fig.~\ref{decayc}. Feeding of $^{140}$Pr$^{58+}$ or
$^{140}$Ce$^{58+}$ ions via radioactive decays or reactions of
other ions has been avoided by blocking the corresponding orbits
in the ESR with mechanical slits.
\par%
Several measurements of the decay of $^{140}$Pr$^{59+}$,
$^{140}$Pr$^{58+}$, and $^{140}$Pr$^{57+}$ ions have been
performed. Decay curves of the parent ions have been fitted with
an exponential function:
\begin{equation}
N_{Pr}(t) = N_{Pr}(0)\cdot e^{-\lambda t},
\end{equation}
where $N_{Pr}(t)$ and $N_{Pr}(0)$ is the number of parent ions at
the time $t$ after injection and at $t=0$, the time of injection,
respectively. For hydrogen-like and helium-like $^{140}$Pr ions,
the decay constant $\lambda$ is the sum of the EC decay constant
$\lambda_{EC}$, the $\beta^+$ decay constant $\lambda_{\beta^+}$,
and the loss constant $\lambda_{loss}$ due to collisions with
residual gas atoms or pick-up of electrons in the electron cooler
($\lambda$ = $\lambda_{EC}$ + $\lambda_{\beta^+}$ +
$\lambda_{loss}$). The bare $^{140}$Pr$^{59+}$ nuclei can only
decay via the $\beta^+$- decay-mode. Hence, the measured decay
constant is the sum $\lambda_{\beta^+}$ + $\lambda_{loss}$. The
growth of the number of daughter ions from the EC decay of
$^{140}$Pr$^{58+}$ into $^{140}$Ce$^{58+}$ nuclei and
$^{140}$Pr$^{57+}$ into $^{140}$Ce$^{57+}$ ions is determined
solely by the EC rate of $^{140}$Pr, whereas the loss of stable
$^{140}$Ce ions is determined only by $\lambda_{loss}$. Therefore,
we can fit the number $N_{Ce}(t)$ of $^{140}$Ce daughters as a
function of time $t$ by using:
\begin{eqnarray}
\nonumber
N_{Ce}(t) = N_{Pr}(0)\cdot
\frac{\lambda_{EC}}{\lambda-\lambda_{loss}} \cdot [
e^{-\lambda_{loss} t} - e^{-\lambda t}] \\
+~ N_{Ce}(0)\cdot e^{-\lambda_{loss} t}
\end{eqnarray}
All measurements have presented consistent results.  The averaged
values for the $\lambda_{EC}$ and $\lambda_{\beta^+}$ decay
constants converted to the rest frame of ions are presented in
Table~\ref{lambdas}. The mean loss constant has been determined to
be $\lambda_{loss} = 0.0003(1)~s^{-1}$, which is within the error
bars the same for the studied charge states of $^{140}$Ce and
$^{140}$Pr.
\par%
As can be seen from Table~\ref{lambdas}, the measured $\beta^+$
decay rate is within the errors independent on the degree of
ionization. This is expected, since the electron screening
modifies the $\beta^+$ rate by less than 3\% in fully-ionized ions
compared to neutral atoms \cite{Dz-Beta}.
\par%
Previously, the EC from the K-orbit has been measured in implanted
atoms by applying X-ray spectroscopy \cite{Ba-RMP}. Such
measurements have been performed for neutral $^{140}$Pr in Refs.
\cite{{Bir-SJP},{Ev-PRC},{Cam-NPA}} and can be compared with our
measurement on the helium-like ions. Using the values for the
$^{140}$Pr$^{57+}$ ions we obtain
$\lambda_{EC}$/$\lambda_{\beta^+}$ = 0.95(8), which agrees well
with 0.90(8) from Ref. \cite{Ev-PRC} and disagrees by about 2.5
standard deviations with 0.74(3) from Ref. \cite{Bir-SJP} and with
0.73(3) from Ref. \cite{Cam-NPA}. We note that it is the first
time that this quantity could be measured directly in helium-like
ions without the influence of other orbital electrons. These
electrons modify the density of the K-electrons at the nucleus by
about 1\% \cite{AI-Priv}, which has been neglected in the above
comparison.
\par%
The striking result is--in spite of the fact that the number of
orbital electrons is {\it reduced} from two in $^{140}$Pr$^{57+}$
ions to only one in $^{140}$Pr$^{58+}$ ions--that the EC-rate {\it
increases} by a factor of 1.49(8). Moreover, the half-life of
$^{140}$Pr$^{58+}$ with a single orbital electron, $T_{1/2}$~=
ln(2)/$\lambda$~= 3.04(9) min, is even shorter than the half-life
$T_{1/2}$~= 3.39(1) min \cite{TOI} of the neutral
$^{140}$Pr$^{0+}$ atoms with 59 orbital electrons.
\begin{table}[!b]
\caption{Measured $\beta^+$ and EC decay constants obtained for
fully-ionized, hydrogen-like, and helium-like $^{140}$Pr ions. The
values are given in the rest frame of the ions.}
\label{lambdas}
\begin{center}
\begin{tabular}{|c|c|c|}
\hline%
Ion&$\lambda_{\beta^+}$ $[s^{-1}]$&$\lambda_{EC}$ $[s^{-1}]$\\
\hline%
$^{140}$Pr$^{59+}$ & 0.00158(8) & --- \\
\hline%
$^{140}$Pr$^{58+}$ & 0.00161(10) & 0.00219(6) \\
\hline%
$^{140}$Pr$^{57+}$ & 0.00154(11) & 0.00147(7) \\
\hline%
\end{tabular}
\end{center}
\end{table}
\par%
Our result can be explained by taking into account the
conservation of total angular momentum of the nucleus-lepton
system. We note, that similar arguments have been used in Refs.
\cite{{Ph-PRA},{Ph-PRL}} to explain the de-excitation of nuclear
excited states decaying via electron conversion in highly-ionized
iron ions and in Ref. \cite{Pr-RMP} to describe the muon-capture
decay rates.
\par
\begin{figure}[t!]
%\begin{center}
\includegraphics*[width=7.5cm]{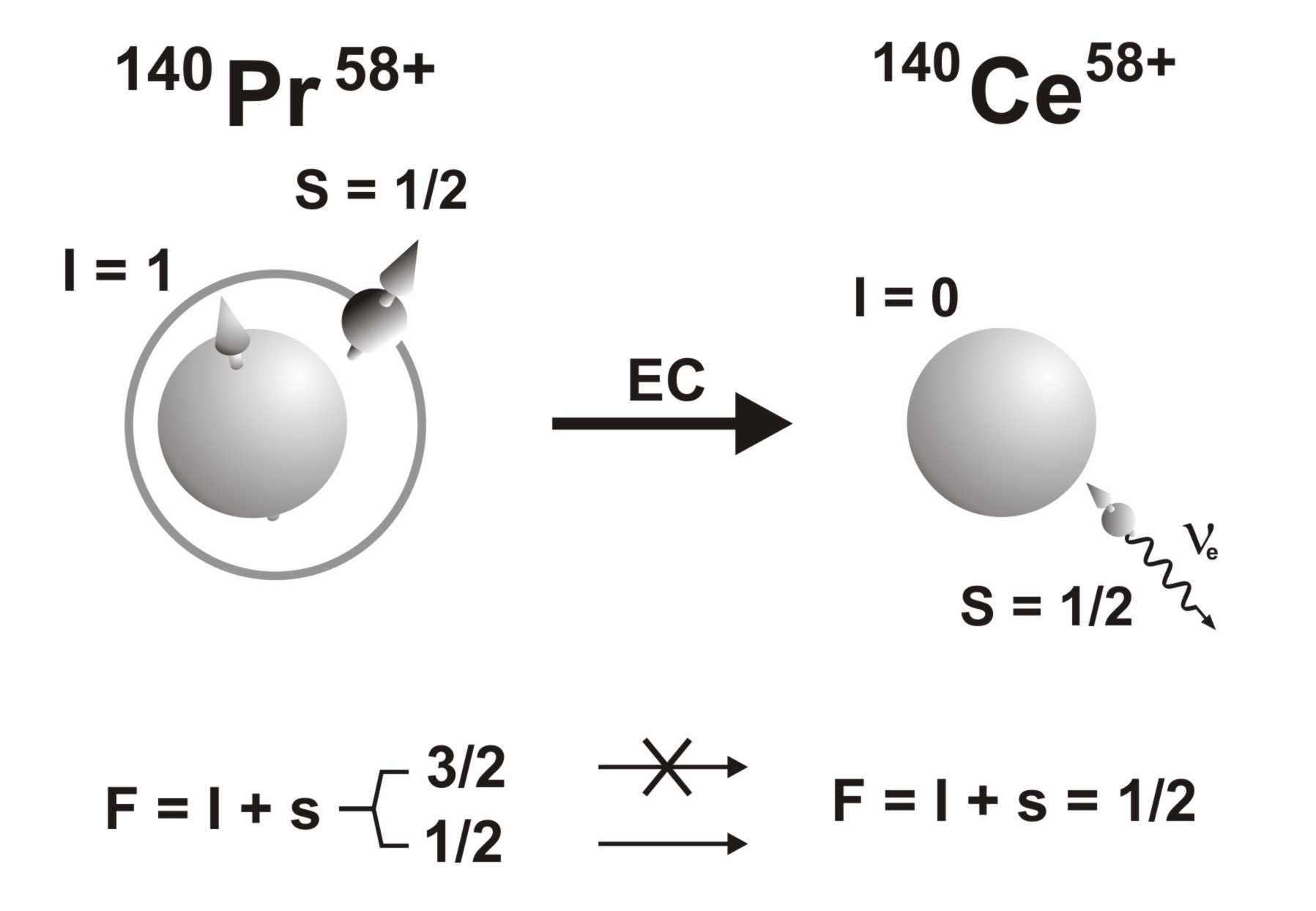}
\caption {Illustration of the EC decay of the hydrogen-like
$^{140}$Pr$^{58+}$ ions to bare $^{140}$Ce$^{58+}$ ions.}
\label{balls}
%\end{center}
\end{figure}

%
In the initial state ($i$), the total angular momentum $F_i$ of a
$^{140}$Pr nucleus with spin $I_i = 1$ and a single bound
K-electron with spin $s = 1/2$ can have two values of the
hyperfine states, $F_i = I_i - s = 1/2$, if the spins of the
nucleus and the electron are anti-parallel, or $F_i = I_i + s =
3/2$ if the spins are parallel, as schematically illustrated in
Fig.~\ref{balls}. In the final state ($f$), however, the total
angular momentum can have only one value, $F_f = 1/2$, which is
the sum of the zero angular momentum of the $^{140}$Ce nucleus
$I_f = 0$ \cite{TOI} and of the spin $s = 1/2$ of the emitted
electron-neutrino. Hence, only transitions from the $F_i = 1/2$
hyperfine state can contribute to the decay to the final state.
The decay from the $F_i = 3/2$ state would require that the
emitted neutrino carries away two units of orbital angular
momentum, which corresponds to a much slower (twice forbidden)
$\beta$-decay.
%
%%%%%%%%%%%%%%%%%%%%%%%%%%%%%%%%%%%%%%%%%%%%%%%%%%%%%%%%%%%%%%%%%%%%%%%%%%%
%%%%%%%%%%%%%%%%%%%%%%%%%%%%%%%%%%%%%%%%%%%%%%%%%%%%%%%%%%%%%%%%%%%%%%%%%%%
\par%
The $F_i = 1/2$ assignment to the lowest hyperfine state of
$^{140}$Pr$^{58+}$ follows from the positive magnetic moment $\mu$
of $^{140}$Pr which has been deduced from the known magnetic
moments of the neighboring odd-A nuclei of about +2.5~$\mu_N$.
%In the present experiment, the ions produced in the excited $F = 3/2$
%state decay via photon emission to the $F = 1/2$ state within a
%time much shorter than the cooling time.
For hydrogen-like $^{140}$Pr, the relaxation time for the upper
hyperfine state to the ground state ($\tau\approx0.03~s$) is much
shorter than the cooling time \cite{Kopf}.
Electric and magnetic fields in the ring can, in principle, lead
to a repopulation of the upper hyperfine level. Such repopulation,
however, has not been observed in ESR experiments \cite{Se-PRL}.
%
%Although electric and magnetic fields--mainly caused by the dipole
%bending magnets of the storage ring--could, in principle, lead to
%a repopulation of the upper hyperfine level, such repopulation has
%not been observed in previous experiments at the ESR
%\cite{Se-PRL}.
%
Thus, $^{140}$Pr$^{58+}$ ions are dominantly stored in a pure $F_i
= 1/2$ quantum state.
\par%
If a nucleus has
$\mu<0$
%a negative magnetic moment
%
than the lower hyperfine state of the hydrogen-like ion is
$F_i=I_i+s$. For instance, for hydrogen-like $^{64}$Cu ions
($\mu=-0.217(2)~\mu_N$, $I_i=1$) \cite{64Cu} the ground state is
$F_i=3/2$ and it does not decay by an allowed EC-decay to the
ground state of $^{64}$Ni ($I_f=0$).
%We note, that if the nucleus has a negative
%magnetic moment, e.g. in the case of $^{64}$Cu
%($\mu=-0.217(2)~\mu_N$, $I_i=1$) \cite{64Cu}, than the lower
%hyperfine state is $F_i=3/2$ and the hyperfine ground state of
%$^{64}$Cu will not decay by an allowed EC-decay to the ground
%state of $^{64}$Ni ($I_f=0$).
\par%
The influence of the hyperfine-state of the electron on the
EC-decay rate at different temperatures has been investigated
theoretically in Ref. \cite{Fo-PRL}, where significant changes in
the decay rates have been predicted. The detailed theoretical
description of our results is given in Ref. \cite{Pa-Th}. This
work provides a systematic study of EC decay rates for
hydrogen-like and helium-like ions dependent on the nuclear spins.
In the case of $^{140}$Pr only one $F_i$-state contributes to the
decay. Then the EC-decay rate depends on the ratio of the
statistical weights of the transition, i. e. $(2I_i + 1)$/$(2F_i +
1)$=3/2, which is in excellent agreement with our experimental
result.
\par%
In summary, our experimental results have clearly revealed a
fundamental property of $\beta$-decay of highly-ionized atoms,
which could not been measured in any previous experiment. The
description of the EC rate which is known from neutral atoms has
to include the conservation of the total angular momentum, in
particular when going to high atomic charge states, which prevail,
e.g., in hot stellar plasmas during nucleosynthesis.
\par%
We would like to acknowledge valuable discussions with
A.N.~Artemyev, W.~Henning, A.N.~Ivanov, K.H.~Langanke,
V.M.~Schabaev, J.~Schiffer, K.~Takahashi, and I.I.~Tupitsyn.
Furthermore we thank the GSI accelerator team for the excellent
technical support.

\end{document}